**Title**

Green Synthesis of Ammonium Nitrate ($NH_4NO_3$) Fertilizer: Production via Plasma Water/Ice Interaction with Air and $NH_3$ Plasma

**Authors**

Vikas Rathore*[1], Vyom Desai[1,2], Nirav I. Jamnapara[1,2], Sudhir Kumar Nema[1,2]

**Affiliations**

[1]Atmospheric Plasma Division, Institute for Plasma Research (IPR), Gandhinagar, Gujarat382428, India

[2]Homi Bhabha National Institute, Training School Complex, Anushaktinagar, Mumbai,400094, India

*Author to whom correspondence should be addressed:*

Vikas Rathore

**\*Email:** vikas.rathore@ipr.res.in, vikasrathore9076@gmail.com

**Abstract**

This study presents a novel and eco-friendly method for synthesizing ammonium nitrate ($NH_4NO_3$) using plasma-activated water (PAW) prepared through air and ammonia ($NH_3$) plasma treatments. Initially, PAW containing nitrate ions ($NO_3^-$) is produced by treating water with air plasma. This PAW (air) is then frozen and exposed to low-pressure $NH_3$ plasma, introducing ammonium ions ($NH_4^+$) to form $NH_4NO_3$ (PAW (air+NH3)). We systematically investigate the voltage-current characteristics of the air and $NH_3$ plasmas, analyze the generated species and radicals to understand the mechanism of $NH_4NO_3$ formation, and evaluate the effects of process parameters such as $NH_3$ gas pressure, applied voltage, and treatment time on the properties of PAW (air+$NH_3$).

Our results indicate that all examined process parameters positively influence the properties of PAW (air+$NH_3$). Among these parameters, the duration of $NH_3$ plasma treatment on PAW (air) ice exerts the most significant effect. Specifically, the concentration of $NH_4^+$ ions increased by 134.2% when the $NH_3$ treatment time was extended from 0.5 hours to 1 hour, compared to 12.7% and 33.3% increases for $NH_3$ gas pressure (ranging from 0.25 to 0.55 mbar) and applied voltage (ranging from 500 to 700 V), respectively. Similarly, variations in pH, oxidation-reduction potential (ORP), and electrical conductivity were substantially higher with increased treatment time than with changes in gas pressure and applied voltage. The PAW (air+$NH_3$) exhibited a neutral to slightly basic pH, making it ideal for soil applications, thereby addressing the existing issue of the high acidity of PAW and its use in agriculture.

These findings highlight a promising green synthesis route for ammonium nitrate, providing an environmentally sustainable alternative to conventional production methods. This approach not only leverages plasma technology for chemical synthesis but also underscores its potential for developing sustainable industrial processes.

*Keywords*: ammonium nitrate, plasma activated water, air plasma, $NH_3$ plasma, reactive oxygen-nitrogen species

## 1. Introduction

Ammonium nitrate ($NH_4NO_3$) is a cornerstone in modern agriculture, serving as a primary nitrogen source for crop growth (1-5). As the global population increases and arable land decreases, there is an urgent need to intensify agricultural production to meet future food demands. It is estimated that by 2050, the world population will reach 9 billion. To feed this large population, advancements in sustainable agricultural technologies, including the eco-friendly and green synthesis of fertilizers, are essential. (6-8). However, traditional methods of synthesizing ammonium nitrate pose significant environmental and safety challenges. To address these issues and meet the rising global demand for efficient fertilizer production, innovative and sustainable approaches are imperative (6, 8-10).

Conventional ammonium nitrate synthesis processes are energy-intensive, leading to high costs and significant carbon emissions. Additionally, these methods often involve hazardous chemicals, posing risks to human health and safety during handling and transportation. The production process also generates greenhouse gases, contributing to air pollution and environmental degradation, and consumes large volumes of water, exacerbating water scarcity issues (7, 8, 11, 12).

Previously, Attri et al. (5) reported the green route for ammonium nitrate synthesis by plasma treatment of soil using $N_2$ as plasma forming gas. The required hydrogen and oxygen for $NH_4^+$ and $NO_3^-$ synthesis were obtained from the moisture present in the $N_2$ gas and the surrounding environment within the vacuum chamber and soil. However, this method limited the production of $NH_4^+$ and $NO_3^-$ ions in the soil. Similarly, the production of ammonium nitrate during the synthesis of $NH_3$ using $N_2$ or air as the plasma-forming gas along with water was limited by the low selectivity of these processes for $NH_4^+$ ion production. These methods often relied on catalytic conversion or electroreduction to enhance selectivity (13-16). Gorbanev et al. (17) reported high selectivity for $NH_4^+$ ion production using a mixture of $N_2$ and water vapor with very low volume of water (5 ml), but the low concentration of $NO_3^-$ ions resulted in significantly low $NH_4NO_3$ production.

To overcome these challenges, the present work introduces a novel, eco-friendly method for ammonium nitrate synthesis. This approach directly converts air and $NH_3$ into $NH_4NO_3$ using plasma-activated water (PAW), generated through plasma treatment of water. By utilizing PAW, this method aims to minimize environmental impact and promote agricultural sustainability. PAW mimics the natural nitrogen fixation process of lightning, where high-energy environments convert atmospheric nitrogen ($N_2$) and oxygen ($O_2$) into reactive nitrogen oxides ($NO_x$), which dissolve in water to form various nitrogen compounds. Unlike the sporadic and uncontrolled nature of lightning, plasma-liquid interaction offers precise control over conditions, improving the efficiency and consistency of nitrogen fixation. This method provides a scalable, environmentally friendly alternative to traditional industrial nitrogen fixation, reducing the environmental impact of chemical fertilizers (18-22).

PAW offers numerous advantages, including antimicrobial properties that can replace chemical pesticides, leading to safer and more sustainable farming practices. Additionally, PAW enhances nutrient absorption in plants, acting as a nitrogen source or fertilizer, resulting in improved crop growth and productivity (23-28).

Central to this synthesis method is the optimization of process parameters such as $NH_3$ gas pressure, applied voltage, and treatment time. Through systematic evaluation, the study aims to identify the most efficient conditions for $NH_4NO_3$ production. Furthermore, the research involves comprehensive characterization of the plasma used in the synthesis process. Voltage-current waveform analysis and emission spectroscopy provide insights into plasma characteristics and reaction mechanisms, facilitating process optimization and understanding.

**2. Materials and methods**

2.1 Production of $NH_4NO_3$ using air plasma and $NH_3$ discharge

The schematic for the production of ammonium nitrate using air plasma and $NH_3$ discharge is shown in Figure 1. In this process, 500 ml of ultrapure Milli-Q water was placed in a 600 ml borosilicate glass beaker. This water was then exposed to air plasma generated by a dielectric barrier discharge (DBD) pencil plasma jet (PPJ). Detailed information about the pencil plasma jet, including its voltage-current characteristics, charge-voltage behavior, and optical emission properties, can be found in our previous studies (29-31).

The plasma-activated water (PAW) produced after air plasma exposure was frozen in a freezer to form rectangular ice cubes, referred to as PAW (air) ice. This PAW (air) ice was subsequently placed in a vacuum chamber for $NH_3$ plasma exposure, as depicted in Figure 1. The vacuum chamber used in this study had dimensions of 250 mm in diameter and 250 mm in height. A rotary vane vacuum pump (HHV pump) with a capacity of 21 m³/h was employed to achieve a base pressure of 0.1 mbar. The chamber was then purged with $NH_3$ gas using a balzer gas dosing valve for precise control (32).

The gas pressure during $NH_3$ insertion was monitored using a vacuum pressure gauge (Pirani gauge model HPH—33 with digital display unit DHPG—011, KMV vacuum technology). The vacuum chamber was electrically grounded (anode), while the cathode was negatively biased. A pulse DC power supply (2.4 kVA, 10 kHz) with a 60% duty cycle was used to power the discharge.

The applied voltage and current across the plasma were measured using a high voltage probe (Tektronix P5100) connected to the cathode and a Rogowski coil (CM-100-MG, Ion Physics Corporation) connected to the anode. A digital oscilloscope (Tektronix TDS 2014C) was used to record these measurements. The optical emission spectrum of the $NH_3$ glow discharge plasma was measured across the wavelength range of 200 to 600 nm using a StellarNet EPP2000-UV spectrometer, with an optical fiber enabling data collection.

2.2 Measurement of physicochemical properties of PAW and RONS concentration

The physicochemical properties of plasma activated water (PAW), including PAW (air) and PAW (air+$NH_3$), were assessed using specialized instrumentation. Electrical conductivity (EC) and total dissolved solids (TDS) were measured using a Contech CC-01 meter and a HM digital AP1 meter, respectively. The pH and oxidation-reduction potential (ORP) were determined using an EUTECH pH Meter (Model: pH700) equipped with a pH Electrode (EC620131) and an ORP electrode (ECFG7960101B). To semi-quantitatively determine the concentration of dissolved ammonium ions ($NH_4^+$), test strips from Macherey-Nagel were utilized, covering ranges of 0 to 6 mg $L^{-1}$ and 0 to 400 mg $L^{-1}$, respectively.

The detection and quantification of reactive oxygen-nitrogen species (RONS) in PAW were conducted using both semi-quantitative and quantitative methods, employing laboratory-grade chemicals and test strips. Reagents such as diazotized sulphanilamide, orthophosphoric acid, sulfuric acid, titanium oxysulfate, and potassium indigo trisulfonate were used. Semi-quantitative analysis involved test strips and colorimetric kits to approximate the concentrations of nitrite ($NO_2^-$), hydrogen peroxide ($H_2O_2$), nitrate ($NO_3^-$) ions, and dissolved ozone ($O_3$) in PAW. Quantitative analysis employed UV-visible spectroscopy, where standard curves for $NO_3^-$, $NO_2^-$, and $H_2O_2$ were prepared using sodium nitrate, sodium nitrite, and a 30% $H_2O_2$ solution, respectively, allowing for precise determination based on the Beer-Lambert law. (19, 25, 26, 33-40).

Specific procedures were followed for detecting individual RONS species. Nitrite ions were detected by reacting them with a reagent to form a reddish-purple azo dye, measurable at 543 nm. Nitrate ions were detected using UV-visible spectroscopy, utilizing their UV absorption at 220 nm. Dissolved ozone was determined via the indigo colorimetric method, relying on the rapid decolorization of the indigo reagent in an acidic solution. Hydrogen peroxide was detected using spectrophotometric analysis with the titanium sulfate method, measuring absorbance at 407 nm (29). To prevent $NO_2^-$ ions interference during $H_2O_2$ analysis, sodium azide ($N_3^-$) was added to PAW to restrict the reaction between nitrite ions and hydrogen peroxide (34, 36).

2.3 Data analysis

All experiments concerning PAW (air) and PAW (air+NH3) were conducted at least three times, and the obtained results were expressed as mean ($\mu$) ± standard deviation ($\sigma$). Statistical analysis was performed with a significance level of 95% (p-value 0.05) using analysis of variance (ANOVA) followed by post-hoc testing with Fisher's Least Significant Difference (LSD) test.

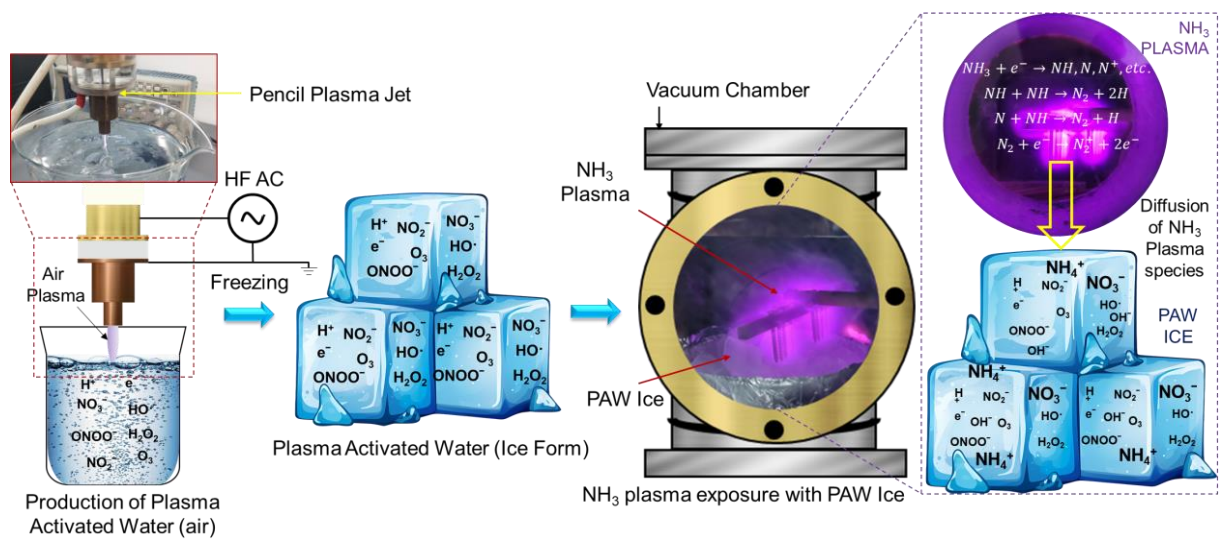

Figure 1. Schematic of production of NH₄NO₃ (aq.) via plasma-water/ice interaction using air and NH₃ plasma.

## 3. Results

3.1 Voltage-current waveform

3.1.1 Air plasma

Figure 2 (a) presents the voltage and discharge current characteristics of the dielectric barrier discharge (DBD) pencil plasma jet (PPJ) operating at atmospheric pressure. The current waveform obtained from this setup exhibits features typical of filamentary DBD micro-discharges, providing valuable insights into the nature and behavior of the plasma generated during air discharge. The DBD plasma consists of numerous filamentary micro-discharges occurring between the dielectric layers. These filamentary micro-discharges are brief, spatially confined discharges resulting from the breakdown of the air in the small gaps between the dielectric barriers. Each filamentary micro-discharge acts as a localized region of ionization, contributing to the overall plasma generation (41).

A sinusoidal voltage (frequency: 40 kHz, peak-to-peak voltage: 10.8 kV) is applied across the dielectric barriers. As the voltage increases, it reaches a threshold where the electric field is strong enough to ionize the air molecules in the gap, leading to the formation of plasma filaments. The periodic nature of the applied voltage ensures that these filamentary micro-discharges occur repeatedly, corresponding to the peaks of the sinusoidal waveform, as shown in Figure 2 (a). The filamentary nature of the DBD plasma is due to localized regions of high electric field that lead to air breakdown. Each filament is a path of ionized gas that conducts the discharge current momentarily. These filaments are not continuous but appear and disappear rapidly, giving rise to the irregular, spiky current waveform observed. This behavior is indicative of the filamentary microcharge nature of DBD cold plasma.

The quartz tube (dielectric) used in the PPJ prevents the formation of a continuous arc, ensuring that the plasma remains in a non-thermal, non-equilibrium state (cold plasma). This is crucial for maintaining the DBD's characteristics and preventing damage to the electrodes. The presence of the dielectric barrier modulates the discharge process, allowing only filamentary micro-discharges to occur and thereby preserving the filamentary nature of the plasma. The discharge power during plasma-water interaction was 23.8 W, calculated using a charge-voltage Lissajous figure (not shown in the manuscript).

3.1.2 $NH_3$ plasma

The voltage and current characteristics of the $NH_3$ plasma discharge at low pressure were analyzed using the oscillograms presented in Figure 2 (b). The observed waveforms provide crucial insights into the discharge dynamics of $NH_3$ gas in a vacuum chamber when subjected to pulse DC voltage. The $NH_3$ discharge was initiated by applying a pulse DC voltage across the electrodes within the vacuum chamber. As the voltage reached its rising peak, the electric field strength increased sufficiently to ionize the $NH_3$ gas molecules, leading to plasma formation. The current waveform exhibited distinct characteristics during the voltage pulses. At the rising peak of the pulse DC voltage, there was a sudden increase in positive current, indicating the onset of NH$_3$ gas discharge. This rise in current corresponds to the formation of plasma as the applied electric field ionizes the NH$_3$ molecules, generating electrons and ions.

As the applied voltage fell steeply, a sudden rise in negative current was observed. This negative current surge indicates the rapid recombination of ions and electrons within the plasma. The decreasing voltage reduces the electric field strength, leading to a decline in ionization rates and an increase in recombination processes. The recombination of ions with electrons results in the emission of photons and the release of energy, observed as a transient spike in the negative current. This periodic behavior suggests that the NH$_3$ plasma discharge occurs predominantly at the rising and falling peaks of the pulse DC voltage. During these peaks, the conditions are most favorable for ionization and subsequent plasma formation, followed by recombination as the voltage decreases.

The low-pressure environment within the vacuum chamber facilitates efficient ionization with minimal collisions between gas molecules, leading to a more uniform and stable plasma. This stability is crucial for applications requiring consistent plasma properties.

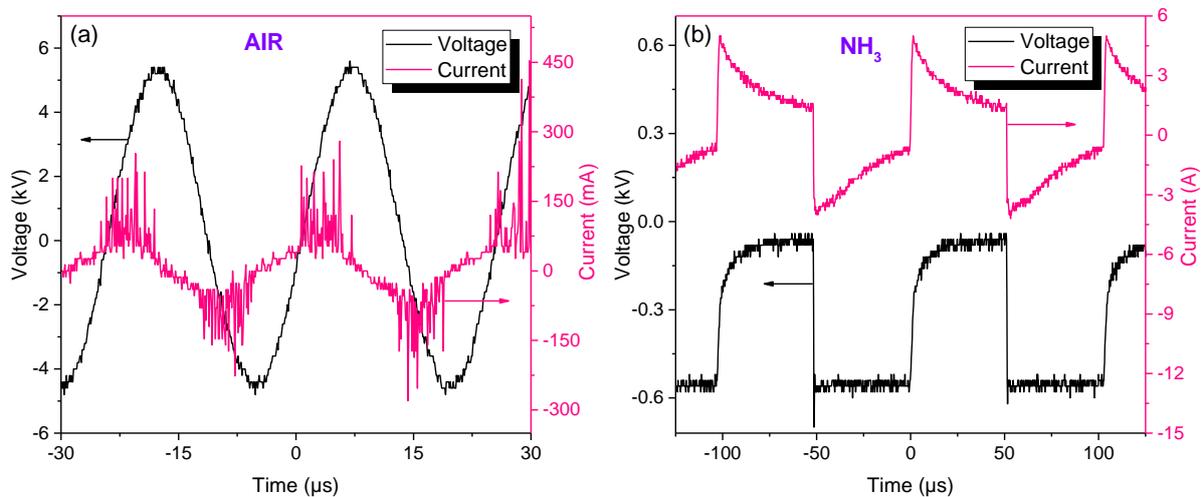

Figure 2. Voltage-current waveform of (a) air and (b) NH₃ plasma produced by pencil plasma jet and low-pressure plasma chamber.

3.2 Species and radicals in air and NH₃ Plasma

3.2.1 Air plasma

The optical emission spectrum of air plasma, shown in Figure 3 (a), provides valuable information about the excited species present. These species undergo electronic transitions from higher to lower energy levels, emitting photons of specific energies. By analyzing these characteristic photon emissions, we can identify the excited species in the plasma. In this study, air was used as the plasma-forming gas to generate plasma-activated water (PAW). The optical emission spectrum reveals various excited species and their corresponding wavelengths.

*Nitrogen Molecule ($N_2$)*

The nitrogen molecule (N₂) in air plasma experiences direct electron impact excitation, transitioning from the ground state (X¹Σg⁺) to the excited state (C³Π_u). The radiative decay (spontaneous emission) from this excited state results in the emission of characteristic photons of the N₂ second positive system (C³Π_u → B³Π_g). The observed vibrational mode transitions (v″, v′) for the N₂ second positive system lies in the range of 312 nm to 435 nm, respectively (42-44).

*Nitrogen Ion (N₂⁺)*

The excitation of N₂⁺ states occur through two primary pathways:

Direct electron impact ionization:

$$N_2(X\,^1\Sigma_g^+)_{v'=0} + e^- \rightarrow N_2^+(B\,^1\Sigma_u^+)_{v''=0} + 2e^- \tag{1}$$

Two-step process:

$$N_2(X\,^1\Sigma_g^+)_{v'=0} + e^- \rightarrow N_2^+(X\,^2\Sigma_g^+)_{v''=0} + 2e^- \tag{2}$$

Followed by:

$$N_2^+(X\,^1\Sigma_g^+)_{v''=0} + e^- \rightarrow N_2^+(B\,^2\Sigma_u^+)_{v''=0} + e^- \tag{3}$$

The radiative decay of N₂⁺ from the excited state (B²Σ_u⁺) to the ground state (X²Σ_g⁺) emits photons in the wavelength range of 390 to 471 nm, respectively (42-44).

*Nitric Oxide (NO)*

Nitric oxide (NO) forms in the plasma phase through reactions between nitrogen and oxygen molecules in the presence of high-energy third bodies (electrons, atoms, or molecules). The radiative decay from the excited state ($A^2\Sigma^+$) to the ground state ($X^2\Pi$) results in weak emission lines at 283 nm (45).

The emission spectrum of air plasma primarily displays strong lines of $N_2$ and traces of N, $N^+$, $N_2^+$, and NO (42-46). The vibrational mode transitions of the $N_2$ second positive system are observed at various wavelengths ranging from 312 nm to 435 nm. The characteristic photons from $N_2^+$ are detected at 391 nm, 419 nm and 470 nm indicating the presence of ionized nitrogen species (42-44). Additionally, lines corresponding to atomic nitrogen (425 nm) and nitrogen ions (399 nm) are present, confirming the dissociation and ionization processes in the plasma. The weak emission line of NO at 283 nm signifies the formation of nitric oxide through high-energy interactions between nitrogen and oxygen molecules.

The optical emission spectrum of air plasma provides a detailed insight into the species present and their corresponding wavelengths. The identification of these species and their transitions is crucial for understanding the underlying plasma chemistry and its applications in generating plasma-activated water. The observed spectral lines confirm the presence of various excited states of nitrogen molecules, nitrogen ions, and nitric oxide, highlighting the complex interactions and reactions occurring in the plasma phase.

3.2.2 $NH_3$ plasma

The emission spectrum of $NH_3$ plasma provides insight into the excited species present during the discharge. Figure 3 (b) shows this spectrum, revealing several key spectral lines and transitions. The following species $N_2^+$, $N_2$, NH, N, $N^+$, etc. were identified. The observed $N_2^+$ vibrational band peaks lie in the wavelength range of 390 to 471 nm, lines correspond to the radiative decay of $N_2^+$ from the excited state ($B^2\Sigma u^+$) to the ground state ($X^2\Sigma g^+$) (44, 46-48).

The $N_2$ vibrational band peaks observed in the wavelength range of 353 and 473, signifies the transition of $N_2$ from the excited state ($C^3\Pi u$) to the ground state ($B^3\Pi g$). Additionally, NH band peaks also observed due to dissociation of $NH_3$ in plasma phase. The presence of NH is indicated by the radiative decay of NH (336 nm) from the excited state ($A^3\Pi_i$) to the ground state ($X^3\Sigma^-$). Along with that, traces of nitrogen atom and ion (N, $N^+$) also observed due to dissociation of NH3 in the wavelength range of 220 to 550 nm, respectively (44, 46-48).

Proposed reactions of $NH_3$ Plasma

The dissociation and subsequent excitation of $NH_3$ molecules under low-pressure conditions drive the reactions of $NH_3$ plasma.

*Dissociation of $NH_3$*

Under the influence of the plasma, $NH_3$ molecules dissociate into various reactive species, primarily N, $N^+$, and NH. This high level of dissociation is facilitated by the low pressure and the presence of high-energy electrons in the plasma (44, 46-48).

*Formation of $N_2^+$ and $N_2$*

$$N + N^+ + M \rightarrow N_2^+ + M \qquad (4)$$

$$N + N + M \rightarrow N_2 + M \qquad (5)$$

Here, '*M*' represents a high-energy third body, which is essential for the stabilization of the newly formed species.

*Radiative Decay Processes:*

$$N_2^+(B\,^2\Sigma_u^+) \rightarrow N_2^+(X\,^2\Sigma_u^+) + h\nu \tag{6}$$

$$N_2(C\,^3\Pi_u) \rightarrow N_2(B\,^3\Pi_g) + h\nu \tag{7}$$

$$NH(A\,^3\Pi_i) \rightarrow NH(X\,^3\Sigma^-) + h\nu \tag{8}$$

*Reactions involving N and $N^+$:*

The presence of N and $N^+$ lines at various wavelengths suggests multiple dissociation and ionization reaction occurs in plasma phase.

$$N_2 + e^- \rightarrow 2N + e^- \tag{9}$$

$$N + e^- \rightarrow N^+ + 2e^- \tag{10}$$

The emission spectrum of $NH_3$ plasma reveals the presence of various excited species, including $N_2^+$, $N_2$, NH, and $N^+$. The characteristic wavelengths of these emissions indicate the specific transitions and radiative decay processes occurring in the plasma. The dissociation of $NH_3$ under low pressure, facilitated by electron impact and subsequent reactions, leads to the formation of these species (44, 46-48). Understanding these mechanisms is crucial for decoding the formation of ammonium nitrate after $NH_3$ plasma exposure with PAW (air) ice.

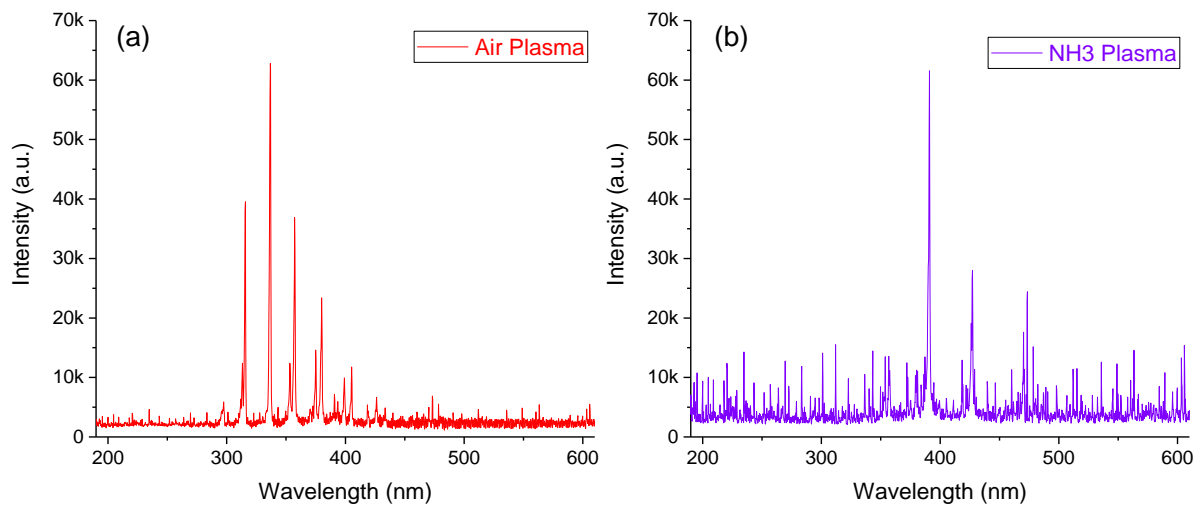

Figure 3. Emission spectra of (a) air and (b) NH$_3$ plasma produced by pencil plasma jet and low-pressure plasma chamber.

3.3 Influence of air plasma on PAW properties and reactive species formation

Air plasma generates high-energy emission bands primarily composed of N$_2$ and N$_2^+$ species and radicals (Figure 3 (a)). These species, in the post-discharge plasma phase, lead to the formation of various gases such as O$_3$, H$_2$O$_2$, and nitrogen oxides (NO$_x$). When these gases dissolve in water, they form stable reactive oxygen-nitrogen species, resulting in significant changes in the physicochemical properties of the water. These changes include a reduction in pH, an increase in oxidizing potential, and a rise in electrical conductivity (23-26, 29, 30, 49-53). Figure 4 presents the measurements of these physicochemical changes and the concentrations of reactive oxygen-nitrogen species in the water.

3.3.1 pH reduction and nitrate/nitrite ions formation

The ultrapure Milli-Q water initially had a pH of 6.95, which decreased to 3.25 after 6 hours of air plasma treatment (Figure 4(a)). This pH reduction is primarily due to the formation of nitrous ($HNO_2$) and nitric acid ($HNO_3$) as $NO_x$ gases dissolve in water, forming $NO_2^-$ and $NO_3^-$ ions (Figure 4(e, f)). The concentration of $NO_3^-$ ions reached 63.5 mg L$^{-1}$, which is over 75 times higher than the $NO_2^-$ ion concentration of 0.84 mg L$^{-1}$. The high concentration of $NO_3^-$ ions in PAW (air) is significant as it provides a crucial source of $NO_3^-$ for the subsequent production of ammonium nitrate ($NH_4NO_3$). The reduction in pH and formation of nitric and nitrous acids after plasma treatment with water have also been shown by Lukes et al. (34) and Lu et al. (36) etc.

3.3.2 Formation of oxidizing species (dissolved $O_3$ and $H_2O_2$)

In addition to acidic species, air plasma also produces oxidizing species such as hydrogen peroxide ($H_2O_2$) and dissolved ozone ($O_3$). $H_2O_2$ is generated through the recombination of hydroxyl radicals ($\cdot OH$), which are produced by the dissociation of moisture ($H_2O$) present in the air during the plasma-liquid interaction. Gaseous $O_3$ is formed in the post-discharge phase when excited oxygen atoms combine with $O_2$ in the air. The maximum concentrations of $H_2O_2$ and dissolved $O_3$ in PAW (air) were 1.51 mg L$^{-1}$ and 2.78 mg L$^{-1}$, respectively (Figure 4 (g, h)). However, at higher plasma treatment durations, the concentration of dissolved $O_3$ declines due to its reaction with $H_2O_2$ and $NO_2^-$ ions, forming more stable $NO_3^-$ ions. This reaction not only decreases the $O_3$ concentration but also reduces the levels of $H_2O_2$ and $NO_2^-$ ions in PAW (air) (34, 36).

3.3.3 Oxidizing Potential

The presence of oxidizing species such as $HNO_3$, $HNO_2$, $H_2O_2$, and dissolved $O_3$ increases the oxidizing potential of water after air plasma treatment (33, 38). After 6 hours of air plasma treatment, the oxidizing potential of ultrapure Milli-Q water increased from 257 mV to 597 mV, representing a 132.3% increase (Figure 4(b)). Ma et al. (33) and Pan et al.(38) also showed an increase in ORP value of water after plasma treatment, which aligns with the results of the present investigation. This increased oxidizing potential, along with the acidic nature of PAW (air), makes it suitable for applications in sterilization, sensitization, and disinfection of various surfaces, medical devices, and food products, etc (23, 24, 27, 38, 52).

3.3.4 Electrical conductivity and total dissolved solids

The generation of inorganic ions ($H^+$, $NO_2^-$, $NO_3^-$) during air plasma treatment increases the total dissolved solids (TDS) and electrical conductivity (EC) of the water (39, 54). After 6 hours of air plasma treatment, the TDS and EC values rose from 0 ppm and 1 µS cm$^{-1}$ to 269 ppm and 684 µS cm$^{-1}$, respectively (Figure 4 (c, d)). These increases in TDS and EC are indicative of the enhanced ion concentration in PAW (air), reflecting its modified physicochemical properties.

Overall, the results demonstrate that air plasma treatment significantly alters the properties of water, creating a highly reactive solution rich in reactive oxygen-nitrogen species. The decrease in pH, increase in oxidizing potential, and rise in electrical conductivity highlight the transformation of ultrapure water into PAW (air) with potential applications in various fields requiring sterilization and disinfection. The formation of $NO_3^-$ ions is particularly noteworthy for its role in the subsequent green synthesis of ammonium nitrate, underscoring the environmental and practical benefits of this plasma-assisted process.

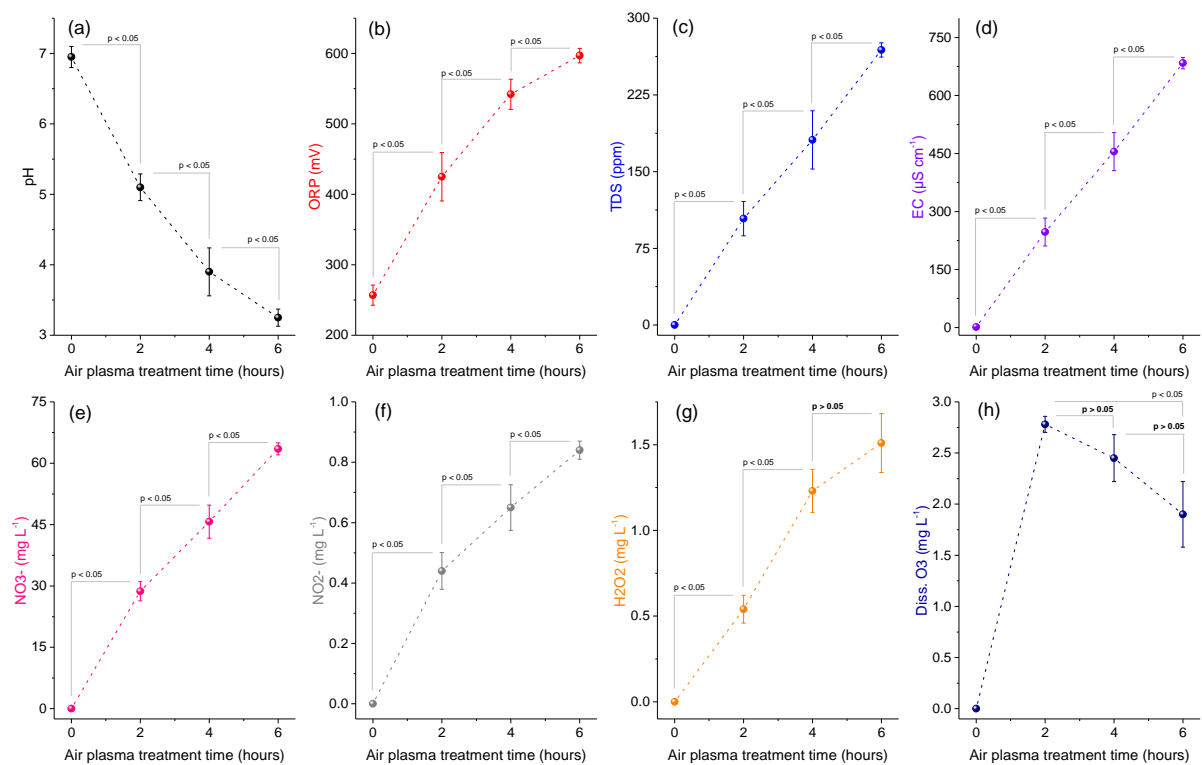

Figure 4. Properties ((a) pH, (b) oxidation-reduction potential (ORP), (c) total dissolved solids (TDS), (d) electrical conductivity (EC), (e) nitrate ($NO_3^-$) ions, (f) nitrite ($NO_2^-$) ions, (g) hydrogen peroxide ($H_2O_2$), and (h) dissolved $O_3$) of plasma-activated water produced by a pencil plasma jet using air as the plasma-forming gas..

3.4 Changes in properties of PAW (air+$NH_3$) and formation of ammonium nitrate

The results shown in Figure 5 illustrate the transformation in the properties of Plasma-Activated Water (PAW+$NH_3$) when air-based PAW ice is subjected to $NH_3$ plasma under sub-atmospheric pressure, leading to the synthesis of ammonium nitrate ($NH_4NO_3$) (13-17). This process holds promise as a green synthesis method for $NH_4NO_3$, which is typically produced through non-environmentally friendly processes. During $NH_3$ plasma exposure, discharge $NH_3$ species and radicals diffuse through PAW ice, forming ammonium ($NH_4^+$) ions. Given that PAW (air) ice already contains nitrate ($NO_3^-$) ions, the presence of $NH_4^+$ ions facilitate the formation of $NH_4NO_3$.

Traditionally, $NH_4NO_3$ is synthesized by reacting ammonium hydroxide ($NH_4OH$) with nitric acid ($HNO_3$) (1-5). In our study, similar reactions were observed (shown in equations (11) and (12)) (47, 55-57), where reactive $NH_3$ species from the plasma formed $NH_4OH$ in PAW (air+$NH_3$) ice, which then reacted with nitric acid present ($HNO_3$) in PAW (air), resulting in the formation of $NH_4NO_3$ (13-17). The effect of various parameters, including $NH_3$ gas pressure, applied voltage, and $NH_3$ plasma treatment time, on the properties of PAW (air+$NH_3$) were systematically analyzed.

$$NH_3(plasma) + H_2O(ice) \rightarrow NH_4OH(ice) \rightarrow NH_4OH(aq.) \quad (11)$$

$$NH_4OH(aq.) + NO_3^- \rightarrow NH_4NO_3(aq.) + HO^- \quad (12)$$

When $NH_3$ plasma interacts with PAW (air) ice, various chemical reactions occur, leading to the formation of ammonium hydroxide ($NH_4OH$) and ammonium nitrate ($NH_4NO_3$).

3.4.1 Proposed Reactions Involved in the Formation of $NH_4NO_3$

*Dissociation of $NH_3$ in plasma:*

Under the influence of plasma, ammonia ($NH_3$) dissociates into nitrogen (N), nitrogen ions ($N^+$), and ammonium radicals (NH) as observed in the emission spectra of $NH_3$ plasma (Figure 3 (b)) (47, 55-57).

$$NH_3 + e^- \rightarrow NH + H_2 + e^- \quad (13)$$

$$2NH_3 + e^- \rightarrow 2N^+ + 3H_2 + 3e^- \quad (14)$$

*Formation of hydroxyl radicals (·OH):*

Hydroxyl radicals are generated in the plasma phase due to the dissociation of water molecules ($H_2O$) evaporated from PAW (air) ice in vapor form.

$$H_2O\ (g) \rightarrow H + \cdot OH \tag{15}$$

*Interaction of $NH_3$ plasma species with PAW (air) ice:*

The species from $NH_3$ plasma (such as NH and $NH_2$) interact with PAW (air) ice, forming $NH_4^+$ (ammonium ions). Moreover, nitric acid ($HNO_3$) in PAW (air) ice reacts with ammonium hydroxide to give ammonium nitrate as a resultant product (47, 55-57).

$$NH + H_2O \rightarrow NH_2 + OH \tag{16}$$

$$NH_2 + H_2O \rightarrow NH_3 + OH \tag{17}$$

$$NH_3 + H_2O \rightarrow NH_4^+ + HO^- \tag{18}$$

$$NH_4^+ + HO^- \rightarrow NH_4OH \tag{19}$$

$$NH_4^+ + NO_3^- \rightarrow NH_4NO_3 \tag{20}$$

These reactions collectively describe the process through which $NH_3$ plasma interacts with PAW (air) ice to form ammonium hydroxide, a process driven by the high-energy environment of the plasma that facilitates the dissociation and recombination of molecules and ions. This ultimately leads to the formation of ammonium nitrate fertilizer in the form of PAW (air+$NH_3$).

3.4.2 Initial observations ($NH_3$ plasma switch off)

Initially, under conditions of 0.25 mbar $NH_3$ gas pressure and a 0.5-hour exposure time without plasma treatment, the PAW (air) exhibited a pH of 3.6, an oxidation-reduction potential (ORP) of 580 mV, an electrical conductivity (EC) of 811 μS cm$^{-1}$, and an $NH_4^+$ ion concentration of 19.3 mg L$^{-1}$. This signifies the absorption of $NH_3$ gas by PAW (air) due to the high solubility of $NH_3$ gas in water, resulting in the observed changes in properties of PAW.

### 3.4.3 pH changes

Exposure to $NH_3$ plasma resulted in an increase in the pH of PAW (air+$NH_3$), attributed to the formation of $NH_4OH$, which neutralized the acidic nature of PAW (air). As shown in Figure 5 (a-c), the pH increase was significant across different treatment parameters. Specifically, increasing the $NH_3$ gas pressure from 0.25 to 0.55 mbar led to an 18.5% increase in pH, which was statistically significant ($p < 0.05$). Similarly, raising the applied voltage from 500 to 700 V resulted in a 19.1% increase in pH. Notably, increasing the $NH_3$ plasma treatment time from 0.5 to 1.0 hours resulted in a 39.7% increase in pH, highlighting the substantial impact of treatment duration on pH levels.

### 3.4.4 ORP changes

The ORP of PAW (air+$NH_3$) decreased as $NH_4OH$, a reducing agent, neutralized the oxidizing species in PAW (air). The trends observed indicated a reduction in ORP with varying treatment parameters. An increase in $NH_3$ gas pressure from 0.25 to 0.55 mbar resulted in a 4% decrease in ORP. Additionally, increasing the applied voltage from 500 to 700 V led to an 8.9% decrease in ORP. The most significant reduction in ORP, 59.5%, was observed when the $NH_3$ treatment time was extended from 0.5 to 1.5 hours, underscoring the critical role of treatment time.

### 3.4.5 Electrical conductivity (EC)

The EC of PAW (air+$NH_3$) increased with $NH_3$ plasma treatment, indicating the formation of more conductive ions, predominantly $NH_4^+$ ions. The data revealed a 15.8% increase in EC with an increase in $NH_3$ gas pressure. The applied voltage increase resulted in a 23.8% rise in EC, while extending the $NH_3$ plasma treatment time led to a significant 152.3% increase in EC. This substantial increase in EC with treatment time suggests an enhanced formation of inorganic ions during the plasma exposure.

### 3.4.6 $NH_4^+$ ion concentration

The concentration of $NH_4^+$ ions was a critical parameter, confirming the formation of $NH_4OH$ and subsequently $NH_4NO_3$. Increasing the $NH_3$ gas pressure from 0.25 to 0.55 mbar resulted in a statistically insignificant ($p > 0.05$) 12.7% increase in $NH_4^+$ ion concentration. However, a significant 33.3% increase was observed when the applied voltage was raised from 500 to 700 V. The most notable increase in $NH_4^+$ ion concentration, reaching 168.2 mg $L^{-1}$, was observed with an $NH_3$ plasma treatment time of 1.5 hours, representing a 134.2% increase from the concentration at a 0.5-hour treatment time. Attri et al. (5) reported a 0.0094 g increase of $NH_4^+$ ions during $N_2$ plasma treatment of soil.

Overall, the findings indicate that $NH_3$ plasma treatment significantly influences the properties of PAW (air+$NH_3$), particularly the $NH_3$ plasma treatment time. This process offers a promising green synthesis method for ammonium nitrate, minimizing the reliance on traditional, non-green synthesis processes. The study's detailed observations of pH, ORP, EC, and $NH_4^+$ ion concentration confirm the feasibility of this approach for environmentally friendly fertilizer production.

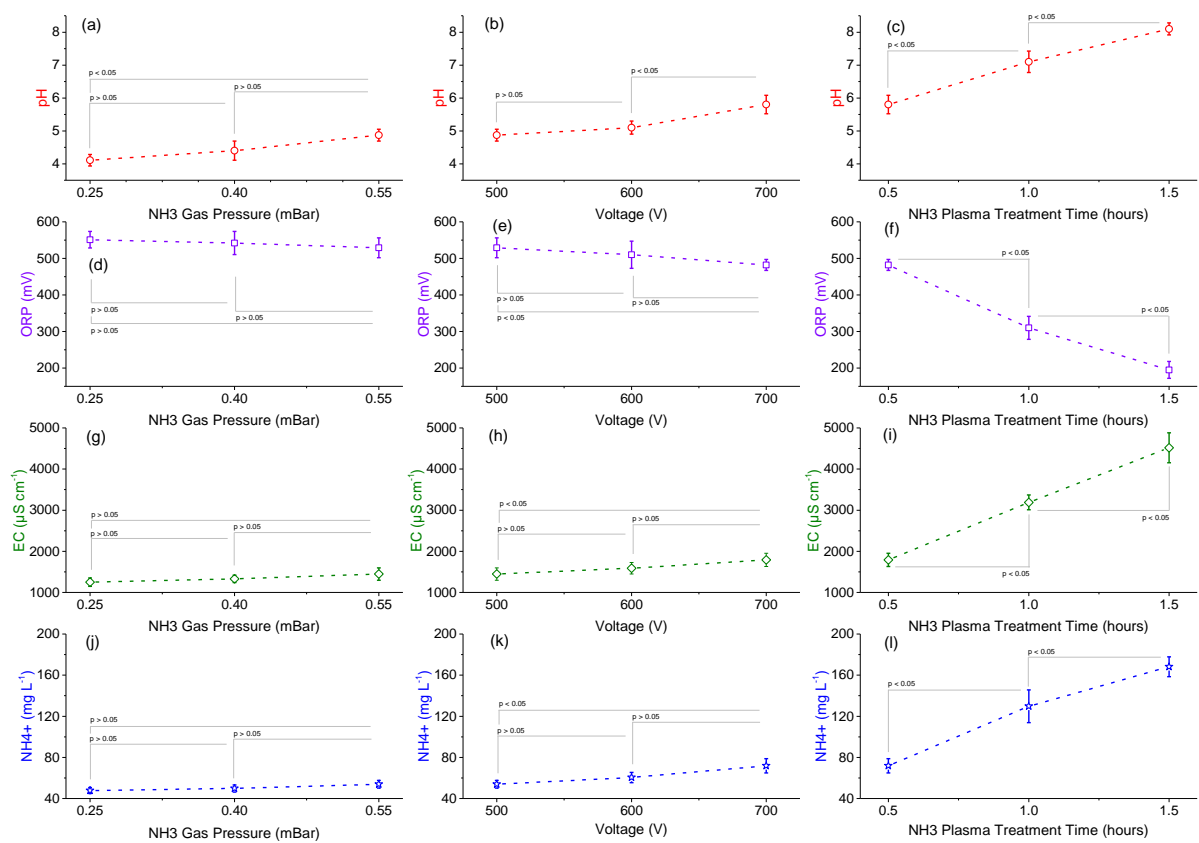

Figure 5. Variation in properties ((a-c) pH, (d-f) ORP, (g-i) EC, and (j-l) $NH_4^+$) of plasma-activated water (air) after exposure with $NH_3$ discharge under varying parameters: gas pressure, applied voltage, and treatment time, leading to the production of ammonium nitrate ($NH_4NO_3$).

## 4. Discussion

This study elucidates the green synthesis of ammonium nitrate ($NH_4NO_3$) via plasma-activated water (PAW), highlighting the interaction between air and ammonia ($NH_3$) plasmas. We provide a comprehensive understanding of the reaction mechanisms underlying $NH_4NO_3$ formation in PAW, emphasizing the role of plasma-induced species and their subsequent interactions in the liquid and ice phase.

Initially, the dielectric barrier discharge (DBD) pencil plasma jet (PPJ) operating at atmospheric pressure generated air plasma, significantly altering the physicochemical properties of ultrapure Milli-Q water. The air plasma, characterized by numerous filamentary micro-discharges, produced a range of reactive species including nitrogen molecules ($N_2$), nitrogen ions ($N_2^+$), nitric oxide (NO), and various excited states of nitrogen atoms (42-46). These species, upon dissolution in water, contributed to the formation of reactive oxygen-nitrogen species such as nitrous acid ($HNO_2$), nitric acid ($HNO_3$), hydrogen peroxide ($H_2O_2$), and dissolved ozone ($O_3$) (26, 33, 34, 36, 38, 58-61). Optical emission spectroscopy confirmed the presence of these species, identifying characteristic photon emissions corresponding to the electronic transitions of these molecules and ions.

Subsequently, exposing this plasma-activated water (PAW) ice to $NH_3$ plasma under sub-atmospheric pressure facilitated the formation of $NH_4NO_3$. The $NH_3$ plasma, generated by applying a pulsed DC voltage, induced the dissociation of $NH_3$ molecules into nitrogen (N), nitrogen ions ($N^+$), and ammonium radicals (NH) (44, 46-48). These reactive species, upon interaction with PAW, engaged in a series of chemical reactions leading to the synthesis of $NH_4NO_3$. The dissociation of $NH_3$ in the plasma phase produced NH and $NH_2$ radicals, which further reacted with hydroxyl radicals (·OH) generated from the dissociation of water molecules in the plasma phase. These hydroxyl radicals played a crucial role in the subsequent formation of ammonium ions ($NH_4^+$) in the aqueous phase.

The presence of nitric acid ($HNO_3$) in PAW, formed during the air plasma treatment, provided the necessary acidic environment for the neutralization reaction with $NH_4OH$, resulting in the formation of $NH_4NO_3$. The interaction of $NH_3$ plasma species with PAW (air) ice, which already contained nitrate ($NO_3^-$) ions, facilitated the recombination of these ions with $NH_4^+$ ions to produce $NH_4NO_3$. The influence of various parameters, including $NH_3$ gas pressure, applied voltage, and $NH_3$ plasma treatment time, was systematically analyzed to understand their impact on the efficiency of $NH_4NO_3$ formation.

Experimental data demonstrated that increasing $NH_3$ gas pressure, applied voltage, and plasma treatment time significantly enhanced the formation of $NH_4^+$ ions and consequently the yield of $NH_4NO_3$. The pH of PAW increased due to the formation of $NH_4OH$, which neutralized the acidic nature of PAW (air). This pH increase was more pronounced with higher $NH_3$ gas pressures, voltages, and extended plasma treatment durations. Correspondingly, the oxidation-reduction potential (ORP) of PAW decreased as $NH_4OH$, a reducing agent, neutralized the oxidizing species present in PAW (air). The electrical conductivity (EC) of PAW increased with $NH_3$ plasma treatment, indicating the formation of more conductive ions, predominantly $NH_4^+$ ions. The concentration of $NH_4^+$ ions increased significantly with higher $NH_3$ gas pressures, voltages, and longer plasma treatment times, confirming the efficiency of this green synthesis approach.

The reaction mechanisms proposed in this study underline the critical role of plasma-generated reactive species and their interactions in the liquid phase. The dissociation of $NH_3$ and water molecules in the plasma phase, followed by the recombination of $NH_4^+$ and $NO_3^-$ ions in the aqueous phase, forms the basis of $NH_4NO_3$ synthesis. The formation of $NH_4OH$ and its subsequent reaction with $HNO_3$ in PAW highlights the intricate balance of plasma chemistry and liquid-phase reactions essential for the efficient synthesis of $NH_4NO_3$.

## 5. Conclusion and future work

This investigation demonstrates the synthesis of ammonium nitrate ($NH_4NO_3$) through plasma-activated water (PAW) prepared using air and $NH_3$ plasma. The study systematically examines the influence of various process parameters, such as $NH_3$ gas pressure, applied voltage, and treatment time, on the properties of PAW (air+$NH_3$), and highlights the potential advantages of this green synthesis route.

A significant finding of this research is the ability of $NH_3$ plasma treatment to neutralize the acidic pH of PAW (air). The initial PAW (air), produced by air plasma treatment of water, exhibited a low pH due to the formation of nitric and nitrous acids. When this PAW (air) ice was exposed to $NH_3$ plasma, the resulting interactions led to the formation of ammonium hydroxide ($NH_4OH$), which neutralized the acidic conditions, raising the pH significantly. This neutralization process is crucial as it facilitates the synthesis of ammonium nitrate by ensuring an optimal reaction environment.

Variations in pH, oxidizing potential, and electrical conductivity were also found to be more significantly affected by $NH_3$ plasma treatment time than by gas pressure or applied voltage. The increase in pH and electrical conductivity, coupled with a decrease in oxidizing potential, underscores the enhanced formation of conductive ions and neutralization of oxidizing species, which are essential for the effective synthesis of $NH_4NO_3$.

In conclusion, the findings of this study demonstrate a promising green synthesis method for ammonium nitrate, leveraging the unique interactions between PAW (air) and $NH_3$ plasma. This method not only offers an environmentally sustainable alternative to traditional ammonium nitrate production processes but also highlights the critical role of $NH_3$ plasma treatment time in optimizing the properties of PAW (air+$NH_3$). These insights pave the way for further advancements in plasma-assisted green synthesis techniques, with potential applications in agriculture and industry.

*Future research will focus on:*

- Scaling Up: Investigating the scalability of the plasma-assisted synthesis method for industrial applications.
- Environmental Impact: Assessing the long-term environmental benefits and potential risks associated with large-scale implementation of this green synthesis approach.

- Application Testing: Evaluating the effectiveness of the synthesized ammonium nitrate in agricultural applications to ensure it meets the required standards for crop production.

By addressing these areas, we can advance the development of sustainable and eco-friendly fertilizer production methods, contributing to global agricultural sustainability and environmental preservation.

**Conflict of interests**

The authors declare that there are no conflicts of interests.